\documentclass[12pt]{article}
\usepackage{a4wide,epsfig,amsmath,amssymb,cite,scalefnt}

\newcommand{\abbrev}{\scalefont{.9}}

\newcommand{\drbar}{$\overline{\mbox{\abbrev DR}}$}
\newcommand{\msbar}{$\overline{\mbox{\abbrev MS}}$}
\newcommand{\asDRbar}{\alpha_s^{\overline{\rm DR}}}
\newcommand{\asMSbar}{\alpha_s^{\overline{\rm MS}}}
\newcommand{\betaDRbar}{\beta^{\overline{\rm DR}}}
\newcommand{\betaMSbar}{\beta^{\overline{\rm MS}}}
\newcommand{\gammaDRbar}{\gamma^{\overline{\rm DR}}}
\newcommand{\gammaMSbar}{\gamma^{\overline{\rm MS}}}
\newcommand{\ZsDRbar}{Z_s^{\overline{\rm DR}}}
\newcommand{\ZsMSbar}{Z_s^{\overline{\rm MS}}}
\newcommand{\ZmDRbar}{Z_m^{\overline{\rm DR}}}
\newcommand{\ZmMSbar}{Z_m^{\overline{\rm MS}}}

\newcommand{\apiDR}{\frac{\asDRbar}{\pi}}
\newcommand{\apiMS}{\frac{\asMSbar}{\pi}}
\newcommand{\aepi}{\frac{\alpha_e}{\pi}}

\newcommand{\qcd}{{\abbrev QCD}}
\newcommand{\susy}{{\abbrev SUSY}}
\newcommand{\dreg}{{\abbrev DREG}}
\newcommand{\dred}{{\abbrev DRED}}

\newcommand{\mDRbar}{m^{\overline{\rm DR}}}
\newcommand{\mMSbar}{m^{\overline{\rm MS}}}

\begin{document}

%- {{{ title and abstract

\title{\vskip-3cm{\baselineskip14pt
    \begin{flushleft}
      \normalsize SFB/CPP-06-08 \\
      \normalsize TTP/06-08 \\
       \normalsize WUB/06-03% --- {\bf v0.13, 24 August 2006}\\
  \end{flushleft}}
  \vskip1.5cm
  Dimensional Reduction applied to QCD at three loops
}
\author{\small R. Harlander$^{(a)}$, P. Kant$^{(b)}$,
  L. Mihaila$^{(b)}$, M. Steinhauser$^{(b)}$\\
  {\small\it (a) Fachbereich C, Theoretische Physik,
    Universit{\"a}t Wuppertal,}\\
  {\small\it 42097 Wuppertal, Germany}\\
  {\small\it (b) Institut f{\"u}r Theoretische Teilchenphysik,
    Universit{\"a}t Karlsruhe,}\\
  {\small\it 76128 Karlsruhe, Germany}\\
}

\date{}

\maketitle

\thispagestyle{empty}

\begin{abstract}
  Dimensional Reduction is applied to \qcd{} in order to compute various
  renormalization constants in the \drbar{} scheme at higher orders in
  perturbation theory. In particular, the $\beta$ function and the
  anomalous dimension of the quark masses are derived to three-loop
  order. Special emphasis is put on the proper
  treatment of the so-called $\varepsilon$-scalars and the additional
  couplings which have to be considered.
\medskip


\end{abstract}

%%%\newpage

%- }}}
%- {{{ Intro:

\section{\label{sec::intro}Introduction}

The big success of the Standard Model of elementary particle
physics is based to a large extent on precision calculations which
sometimes reach three-, four- and even five-loop accuracy.
Such calculations currently all rely on
Dimensional Regularization (\dreg{})~\cite{'tHooft:1972fi,Bollini:1972ui} 
which is an elegant and 
powerful tool to parametrize the divergences occurring at intermediate
steps of the calculations. 

In \dreg{}, the number of space-time dimensions is altered from four to
$D=4-2\epsilon$, which renders the loop integrations finite.  It is
clear, however, that if \dreg{} is applied to a 4-dimensional
supersymmetric theory, the number of bosonic and fermionic degrees of
freedom in super-multiplets is no longer equal, such that supersymmetry
(\susy{}) is explicitly broken.  In order to avoid this problem,
Dimensional Reduction (\dred{}) has been suggested as an alternative
regularization method~\cite{DRED}. Space-time is compactified to
$D=4-2\epsilon$ dimensions in \dred{}, such that the number of vector
field components remains equal to four. Momentum integrations are
$D$-dimensional, however, and divergences are parametrized in terms of
$1/\epsilon$ poles, just like in \dreg{}. Since it is assumed that
$\epsilon> 0$, the four-dimensional vector fields can be decomposed in
terms of $D$-dimensional ones plus so-called $\varepsilon$-scalars.  The
occurrence of these $\varepsilon$-scalars is therefore the only
difference between \dreg{} and \dred{}, so that all the calculational
techniques developed for \dreg{} are applicable also in \dred{}.

Nevertheless, it soon was realized that \dred{} suffers from
mathematical inconsistencies in its original
formulation~\cite{Siegel:1980qs}. Currently, it seems that they can only
be avoided by interpreting the fields as living in an infinite
dimensional space, which again leads to explicit \susy{}
breaking~\cite{Avdeev:1981vf,Stockinger:2005gx}.  A higher order
calculation will therefore require similar \susy{} restoring counter
terms as they are needed in \dreg{} in general.  For some of the
currently available two-loop results, however, it has been shown that
these counter terms vanish~\cite{Hollik:2005nn}.

Although \dred{} was originally constructed for applications in
supersymmetric models, it has been shown that in certain cases it can be
useful also in non-supersymmetric
theories~\cite{Jack:1993ws,Smith:2004ck,Signer:2005iu} like \qcd{}.  For
example, since it is possible to turn \qcd{} (with massless quarks) into
a super-Yang-Mills theory by simply adjusting the colour factors, a
calculation using \dred{} provides the possibility to use non-trivial
Ward identities for a check of complicated calculations (see, e.g.,
Ref.~\cite{Kunszt:1993sd}).

As mentioned before, \dred{} parametrizes ultra-violet divergences as
poles in $\epsilon$. One can therefore formulate a renormalization
scheme analogous to the \msbar{} scheme, usually called the \drbar{}
scheme. In this paper, we compute the beta function of the strong
coupling and the anomalous dimension of the quark masses to three-loop
accuracy within this scheme.  An important issue turns out
to be the renormalization of the $q\bar q\varepsilon$ vertex. It
requires to introduce a new, so-called evanescent coupling constant
$\alpha_e$. A similar argument holds for the four-$\varepsilon$-scalar
vertex, but at the order considered here, this vertex does not get
renormalized.
The proper treatment of $\alpha_e$ leads us to conclude
that the three-loop result for the \qcd{} $\beta$ function available in
the literature~\cite{Bern:2002zk} is incorrect. The correct result is
provided in Section~\ref{sec::beta}.

The outline of the paper is as follows. In Section~\ref{sec::frame} we
provide the notation and set the general framework for the
calculation. Subsequently, we describe in Sections~\ref{sec::beta}
and~\ref{sec::gamma} the calculation for the $\beta$ and the $\gamma_m$
function up to three loops. Section~\ref{sec::concl} contains
the conclusions.

%- }}}
%- {{{ Framework

\section{\label{sec::frame}Framework}

We consider \qcd{} and apply Dimensional Reduction (\dred{}) as the
regularization scheme.  Thus, besides the usual \qcd{} Feynman rules for
quarks ($q$) and gluons ($g$), we have to consider additional vertices
involving the so-called $\varepsilon$-scalars, namely
$q\bar{q}\varepsilon, g\varepsilon\varepsilon, gg\varepsilon\varepsilon,
\varepsilon\varepsilon\varepsilon\varepsilon$ (for the corresponding
Lagrange density, see, e.g.,
Refs.~\cite{Capper:1979ns,Bednyakov:2002sf}).  In general, also a mass
term for the $\varepsilon$-scalar has to be taken into
account~\cite{Jack:1994kd,Jack:1994rk}. However, on simple dimensional
grounds it affects neither the \qcd{} $\beta$-function nor the anomalous
dimension of the quark mass, so we do not need to consider it here.

In a non-supersymmetric theory, it is important to note that the $q\bar
q\varepsilon$ and the $q\bar qg$ vertices renormalize
differently. Therefore, one needs to distinguish the coupling constant
$g_e$, multiplying the $q\bar q\varepsilon$ vertex, from the strong
coupling $g_s$~\cite{Jack:1993ws}. Also the
$\varepsilon\varepsilon\varepsilon\varepsilon$ vertex renormalizes
differently; in fact, in \qcd{} one needs to allow for a more general
colour structure of this vertex, leading to three additional coupling
constants $\lambda_r$ ($r=1,2,3$).  In order to fix the notation we
display the relevant part of the Lagrange density~\cite{Jack:1993ws}
\begin{eqnarray}
  {\cal L} &=& \ldots 
  -\frac{1}{4}\sum_{r=1}^3\lambda_r
  H_r^{abcd}
   \varepsilon_\sigma^a  \varepsilon_{\sigma^\prime}^c
  \varepsilon_\sigma^b  \varepsilon_{\sigma^\prime}^d 
  + \ldots
  \,,
\end{eqnarray}
where  $ \varepsilon$ denotes the $\varepsilon$-scalar fields, and $\sigma$ and
$\sigma^\prime$ 
are $2 \varepsilon$-dimensional indices. For the $SU(3)$ gauge group, the
 $H^{abcd}_r$ are three
independent rank four tensors  which are 
symmetric under the interchange of $(ab)$ and $(cd)$.
Our choice
\begin{eqnarray}
  H^{abcd}_1 &=& \frac{1}{2}\left (f^{ace} f^{bde}  + f^{ade} f^{bce}\right) 
    \,,\nonumber\\
  H^{abcd}_2 &=& \delta^{ab} \delta^{cd} 
  + \delta^{ac} \delta^{bd}  + \delta^{ad} \delta^{bc} 
  \,,\nonumber\\
  H^{abcd}_3 &=&  \frac{1}{2}\left(
  \delta^{ac} \delta^{bd} + \delta^{ad} \delta^{bc}\right)
  - \delta^{ab} \delta^{cd}
  \,,
\end{eqnarray}
fixes the Feynman rules in a unique way. Note that for $SU(N_c), N_c>3$, 
there are four independent tensors $H_r^{abcd}$.
At the order considered in this paper no renormalization constant for
the $\varepsilon\varepsilon\varepsilon\varepsilon$ vertex has to be
introduced.
The vertices $g\varepsilon\varepsilon$ and
$gg\varepsilon\varepsilon$, on the other hand, are renormalized
according to $g_s$ because of gauge invariance~\cite{Jack:1993ws}.

$g_e$ and $\lambda_r$ will be called ``evanescent couplings'' in what
follows, and we define
\begin{eqnarray}
  \alpha_s = \frac{g_s^2}{4\pi}\,,
  \quad
  \alpha_e = \frac{g_e^2}{4\pi}
  \quad\mbox{and}\quad 
  \eta_r = \frac{\lambda_r}{4\pi}
  \,.
\end{eqnarray}

The renormalization constants for the couplings $g_s$ and $g_e$, the quark
mass $m$, the \qcd{} gauge parameter $\xi$, as well as for the fields
and the vertices are introduced as
\begin{align}
  g_s^0 &= \mu^{\epsilon}Z_s g_s\,,\qquad &
  g_e^0   &= \mu^{\epsilon}Z_e g_e\,,\qquad &
  m^0   &= m Z_m\,,\qquad &
  \nonumber\\
  1-\xi^0 &= \left(1-\xi\right) Z_3\,,\qquad &
  q^{0} &= \sqrt{Z_2}\,q\,,\qquad &
  G_\mu^{0,a}    &= \sqrt{Z_3}\,G_\mu^a\,,\qquad &
  \nonumber\\
  \varepsilon^{0,a}_\sigma &=
  \sqrt{Z_3^\varepsilon}\,\varepsilon^a_\sigma\,, \qquad & 
  c^{0,a}        &= \sqrt{\tilde{Z}_3}\,c^a\,,\qquad &
  \bar c^{0,a}        &= \sqrt{\tilde{Z}_3}\,\bar c^a\,,\qquad &
  \nonumber\\
  \Gamma_{q\bar{q}G}^{0}        &= Z_1 \Gamma_{q\bar{q}G}\,, \qquad &
  \Gamma_{q\bar{q}\varepsilon}^{0} &= Z_1^\varepsilon
  \Gamma_{q\bar{q}\varepsilon}\,, 
  \qquad &
  \Gamma_{c\bar{c}G}^{0}              &= \tilde{Z}_1 \Gamma_{c\bar{c}G}\,,
  \label{eq::renconst}  
\end{align}
where $\mu$ is the renormalization scale, $D=4-2\epsilon$ is the number
of space-time dimensions, and the bare quantities are marked by the
superscript ``0''. The quark, gluon, $\varepsilon$-scalar, and ghost
fields are denoted by $q$, $G^a_\mu$, $\varepsilon^a$ and $c^a$,
respectively, and $\Gamma_{xyz}$ stands for the vertex functions
involving the particles $x$, $y$ and $z$ ($a$ is the colour index.).
The gauge parameter $\xi$ is defined through the gluon propagator,
\begin{eqnarray}
  D_g^{\mu\nu}(q) &=& -i\,
  \frac{g^{\mu\nu}- \xi \frac{q^\mu q^\nu}{q^2}}
  {q^2+ i\varepsilon}
  \,.
\end{eqnarray}

From the renormalization of the ghost-gluon or quark-gluon vertex 
one obtains the renormalization constant of the strong coupling
\begin{eqnarray}
  Z_s &=& \frac{\tilde{Z}_1}{\tilde Z_3\sqrt{Z_3}}
  \,\, = \,\, \frac{Z_1}{Z_2\sqrt{Z_3}}
  \,.
  \label{eq::Zg}
\end{eqnarray}
Similarly, the quark--$\varepsilon$-scalar vertex leads to the relation
\begin{eqnarray}
  Z_e &=& \frac{Z_1^\varepsilon}{Z_2\sqrt{Z_3^\varepsilon}}
  \,.
  \label{eq::Zh}
\end{eqnarray}
It is well-known that $Z_s\not= Z_e$ even at one-loop order.
Furthermore, both $Z_s$ and $Z_e$ depend on $g_s$, 
$g_e$, and $\lambda_r$~\cite{Jack:1993ws}; 
note, however, that $Z_s$ depends on $g_e$ and $\lambda_r$
only starting from three- and four-loop order, respectively, 
while $Z_e$ depends on $g_e$ and  $\lambda_r$ already at 
one- and two-loop order, respectively. 

Let us next introduce the $\beta$ functions both for \dreg{} and \dred{}.
In \dreg{}, of course, the $\varepsilon$-scalars are absent,
and from the definition
\begin{eqnarray}
   \betaMSbar(\asMSbar)
   &=& \mu^2 \frac{{\rm d}}{{\rm d}\mu^2} \apiMS
   \label{eq::beta_DREG}
\end{eqnarray}
the usual relation between $\betaMSbar$ and $Z_s$ is obtained:
\begin{eqnarray}
  \betaMSbar(\asMSbar)
  &=& -\epsilon \apiMS
  \left(1+ 2 \asMSbar\frac{\partial \ln \ZsMSbar }{\partial
  \asMSbar} \right)^{-1} 
   \label{eq::beta_DREG_2}
  \,,
\end{eqnarray}
where $\ZsMSbar$ denotes $Z_s$ evaluated in the \msbar{} scheme,
and $\asMSbar$ is the usual definition of the strong coupling
within this scheme~\cite{Bardeen:1978yd}. $\betaMSbar$ is known to four-loop
order (see Refs.~\cite{vanRitbergen:1997va,Czakon:2004bu} and references
therein).  Due to the fact that we have five different couplings in
\dred{}, the relations between $Z_s$ and $Z_e$ and the corresponding
beta functions are slightly more involved. They are given by
\begin{eqnarray}
  \lefteqn{ \betaDRbar_s(\asDRbar,\alpha_e, \{\eta_r\})
 = \mu^2 \frac{{\rm d}}{{\rm d}\mu^2} \apiDR}
 \nonumber\\
  &=& - \left(\epsilon \apiDR + 2 \frac{\asDRbar}{\ZsDRbar} 
    \frac{\partial \ZsDRbar}{\partial \alpha_e} \beta_e
    + 2 \frac{\asDRbar}{\ZsDRbar}\sum_r 
    \frac{\partial \ZsDRbar}{ \partial \eta_r} \beta_{\eta_r} 
  \right)
  \left(1+  2 \frac{\asDRbar}{\ZsDRbar} \frac{\partial \ZsDRbar}
    {\partial \asDRbar}\right)^{-1}    
  \,,\nonumber\\
   \lefteqn{\beta_e(\asDRbar,\alpha_e, \{\eta_r\})
  = \mu^2 \frac{{\rm d}}{{\rm d}\mu^2} \aepi}
 \nonumber\\
  &=& - \left(\epsilon \aepi + 2 \frac{\alpha_e}{Z_e} 
    \frac{\partial Z_e}{\partial \asDRbar} \betaDRbar_s
    + 2 \frac{\alpha_e}{Z_e} \sum_r 
    \frac{\partial Z_e}{\partial \eta_r} \beta_{\eta_r}
  \right) 
  \left(1+  2 \frac{\alpha_e}{Z_e} \frac{\partial Z_e}
    {\partial \alpha_e}\right)^{-1} 
  \, ,
  \label{eq::Zg_beta}
\end{eqnarray}
where it is understood that the renormalization constants $Z_e$ and
$\ZsDRbar$ in Eq.~(\ref{eq::Zg_beta}) are evaluated within \dred{} with
(modified) minimal subtraction, and $\asDRbar$ is the corresponding
strong coupling constant in this scheme. 
As in the \msbar{} scheme, the
coefficients of the single poles fully determine the $\beta$ functions.
%In Eq.~(\ref{eq::Zg_beta})  the dots indicate that terms depending
%on higher orders in the quartic $\varepsilon$-scalar couplings
%$\lambda_r$ have been dropped, cf.~Section~\ref{sec::frame}.  
%In the applications discussed in this paper not all contributions to 
%the expressions of Eq.~(\ref{eq::Zg_beta}) are needed. In particular,
%the terms proportional to $\beta_{\eta_r}$, the beta functions
%corresponding to the couplings $\eta_r$, contribute to
%$\betaDRbar_s$ only  at the four-loop order. Furthermore, the
%approximation $\beta_{\eta_r} = -\epsilon \frac{\eta_r}{\pi}$
%is needed for the two-loop calculation of $\beta_e$.
Let us remark that the terms proportional to $\beta_{\eta_r}$,
the beta functions corresponding to the couplings $\eta_r$, contribute to
$\betaDRbar_s$ only at the four-loop order. Furthermore, only the
approximation $\beta_{\eta_r} = -\epsilon \frac{\eta_r}{\pi}$
is needed for the two-loop calculation of $\beta_e$.

In analogy to Eqs.~(\ref{eq::beta_DREG_2}) and~(\ref{eq::Zg_beta}) we
introduce the anomalous mass dimensions which are given by
\begin{eqnarray}
   \gammaMSbar_m(\asMSbar)
  &=& \frac{\mu^2}{\mMSbar} \frac{{\rm d}}{{\rm d}\mu^2} \mMSbar
  \,\,=\,\, - \pi\betaMSbar
  \frac{\partial \ln \ZmMSbar}{\partial \asMSbar}\,,
  \nonumber\\
  \gammaDRbar_m(\asDRbar,\alpha_e, \{\eta_r\})
  &=& \frac{\mu^2}{\mDRbar} \frac{{\rm d}}{{\rm d}\mu^2} \mDRbar
 \nonumber\\
  &=&
   - \pi \betaDRbar_s
  \frac{\partial \ln \ZmDRbar}{\partial \asDRbar}
  -  \pi \beta_e \frac{\partial  \ln \ZmDRbar }{\partial \alpha_e} 
  -\pi \sum_r \beta_{\eta_r} \frac{\partial  \ln \ZmDRbar
  }{\partial \eta_r}
  \,.
  \label{eq::Zm_gamma}
\end{eqnarray}
As in the case of the $\beta$ function, $\gammaDRbar_m$ also gets
additional terms due to the dependence of $Z_m$ on 
the evanescent coupling $g_e$ and on the quartic $\varepsilon$-scalar
couplings $\lambda_r$ .  
The four-loop result for $\gammaMSbar_m$ can be found in
Refs.~\cite{Chetyrkin:1997dh,Vermaseren:1997fq}.

Let us add a few remarks concerning the meaning of the evanescent
coupling $\alpha_e$ at this point.  In a non-supersymmetric theory,
$\alpha_e$ can be set to an arbitrary value $\hat \alpha_e$ at an
arbitrary, fixed scale $\hat\mu$, $\alpha_e(\hat \mu)\equiv
\hat\alpha_e$. This corresponds to a choice of scheme and in turn
determines the value of $\asDRbar$ through, say, an experimental
measurement. At any scale $\mu$, both $\asDRbar$ and $\alpha_e$
are then determined by the renormalization group
equations~(\ref{eq::Zg_beta}). One particular scheme choice would be to
set $\alpha_e(\hat\mu) = \asDRbar(\hat \mu)$. Note, however, that
already at one-loop level one will have $\alpha_e(\mu)\neq
\asDRbar(\mu)$ for any $\mu\neq \hat\mu$ due to the difference in the
renormalization group functions $\beta_s$ and $\beta_e$.

In a supersymmetric theory, on the other hand, one necessarily has
$\betaDRbar_s=\beta_e$ and $\asDRbar=\alpha_e$ at all scales. Thus,
if one assumes that \qcd{} is a low energy effective theory of \susy{}-\qcd{},
$\alpha_e$ is no longer a free parameter. Rather, $\alpha_e$ and
$\asDRbar$ are both related to the unique \susy{}-\qcd{} gauge coupling by
matching relations (see, e.g., Ref.\,\cite{Harlander:2005wm}) and
renormalization group equations.

These considerations show that the choice $\alpha_e=\asDRbar$ 
is not compatible with the
renormalization group evolution of these couplings unless all \susy{}
particles are taken into account in the running. In fact, it cannot be
assumed at {\it any} scale as soon as one or more \susy{} particles are
integrated out.  An example where this is relevant already at one-loop
level is the $\mDRbar\leftrightarrow\mMSbar$ relation as will be pointed
out in connection with Eq.\,(\ref{eq::mDRmMS}) below.
An analogous discussion holds also for the evanescent couplings $\eta_r$.

%- }}}
%- {{{ beta function to 3-loop order:

\section{\label{sec::beta}$\beta$ function to three-loop order}

Within the framework of \dred{} outlined in the previous section we 
have computed $Z_1$, $Z_2$, $Z_3$, $\tilde{Z}_1$ and $\tilde Z_3$
to three-loop order. They are obtained from the two- and three-point
functions according to Eq.~(\ref{eq::renconst}) (see, e.g.,
Ref.~\cite{Steinhauser:1998cm} for explicit formulae).
Thus, according to Eq.~(\ref{eq::Zg}), $Z_s$ is computed in two
different ways and complete agreement is found.
Furthermore, we compute $Z_1^\varepsilon$ and $Z_3^\varepsilon$ to
two-loop order and hence obtain $Z_e$ to the same approximation.  

Since only the divergent parts enter the renormalization constants, we
can set all particle masses to zero and choose one proper external
momentum in order to avoid infrared problems. For the generation of the
about 11,000 diagrams we use {\tt QGRAF}~\cite{Nogueira:1991ex} and
process the diagrams with {\tt q2e} and {\tt
exp}~\cite{Seidensticker:1999bb,Harlander:1997zb} in order to map them
to {\tt MINCER}~\cite{Larin:1991fz} which can compute massless one-,
two- and three-loop propagator-type diagrams.

The $n$-loop calculation leads to counter terms for $g_s$, $g_e$, and the
gauge parameter $\xi$, which are then inserted into the $(n+1)$-loop
calculation in order to subtract the sub-divergences.  We remark that
$\varepsilon$-scalars are treated just like physical particles in this
procedure.

For the $\beta$ function, to a large extent it is possible to avoid the
calculation with 
$\varepsilon$-scalars and evaluate the Feynman diagrams by applying only
slight modifications as compared to \dreg{}. For that, after the
projectors have been applied and the traces have been taken in
$D=4-2\epsilon$ dimensions, one sets $\epsilon=0$.  The evaluation of
the momentum integrals, however, proceeds in $D$ dimensions, just as for
\dreg{}. During the calculation it is necessary to keep track of the
$q\bar{q}g$ vertices since the difference between \dreg{} and \dred{} in
the results of the corresponding diagrams effectively accounts for the
contributions from the $q\bar{q}\varepsilon$ vertex. Thus, the
renormalization constant $Z_e$ has to be used for this contribution.

We refrain from listing explicit results for $Z_s$ but instead present
the results obtained from Eq.~(\ref{eq::Zg_beta}).  Although
$\beta_e$ is only needed to one-loop order
for the three-loop calculation of 
$\betaDRbar_s$ we present the two-loop expression which enters
the three-loop calculation of $\gammaDRbar_m$.
Writing
\begin{equation}
  \begin{split}
    \betaDRbar_s(\asDRbar,\alpha_e,  \{\eta_r\}) &=
    - \epsilon\apiDR - \sum_{i,j,k,l,m} \betaDRbar_{ijklm}
    \left(\apiDR\right)^i\left(\aepi\right)^j
    \left(\frac{\eta_1}{\pi}\right)^k
    \left(\frac{\eta_2}{\pi}\right)^l 
    \left(\frac{\eta_3}{\pi}\right)^m 
% + \ldots
    \,,\\
    \beta_e(\asDRbar,\alpha_e, \{\eta_r\}) &=
    - \epsilon\aepi
    - \sum_{i,j,k,l,m} \beta^{e}_{ijklm} 
    \left(\apiDR\right)^i
    \left(\aepi\right)^j 
    \left(\frac{\eta_1}{\pi}\right)^k
    \left(\frac{\eta_2}{\pi}\right)^l 
    \left(\frac{\eta_3}{\pi}\right)^m 
    \,,
  \end{split}
\label{eq::defbetaDR}
\end{equation}
we find for the non-vanishing
coefficients up to three respectively two loops:
\begin{eqnarray}
        \betaDRbar_{20}
        &=& \frac{11}{12}C_A - \frac{1}{3} T n_f \,,
        \nonumber\\
        \betaDRbar_{30}
        &=&
        \frac{17}{24} C_A^2 
        - \frac{5}{12} C_A T n_f 
        - \frac{1}{4} C_F T n_f\,,
        \nonumber\\
        \betaDRbar_{40}
        &=&
    \frac{3115}{3456} C_A^3 
  - \frac{1439}{1728} C_A^2 T n_f  
  - \frac{193}{576} C_A C_F T n_f \nonumber \\ 
  & &  + \frac{1}{32} C_F^2 T n_f  
  + \frac{79}{864} C_A T^2 n_f^2 
  + \frac{11}{144} C_F T^2 n_f^2 
        \,,\nonumber \\
        \betaDRbar_{31}
        &=&
  - \frac{3}{16} C_F^2 T n_f
        \,,\nonumber \\
        \betaDRbar_{22}
        &=&
  - C_F T n_f \left(
  \frac{1}{16} C_A - \frac{1}{8} C_F  
  - \frac{1}{16} T n_f \right) 
  \,,\nonumber\\
        \beta_{02}^e
        &=&
  - C_F - \frac{1}{2} T n_f + \frac{1}{2} C_A
        \,,\nonumber\\
        \beta_{11}^e
        &=&
        \frac{3}{2}C_F 
        \,,\nonumber\\
        \beta_{03}^e
        &=&
        \frac{3}{8} C_A^2 - \frac{5}{4} C_A
        C_F + C_F^2 - \frac{3}{8} C_A T n_f + \frac{3}{4} C_F T n_f
        \,,\nonumber\\
        \beta_{21}^e
        &=&
        - \frac{7}{64} C_A^2 + \frac{55}{48}  C_A C_F + \frac{3}{16} C_F^2 
        + \frac{1}{8} C_A T n_f - \frac{5}{12} C_F T n_f 
        \,,\nonumber\\
        \beta_{12}^e
        &=& - \frac{3}{8} C_A^2 + \frac{5}{2} C_A C_F 
        - \frac{11}{4} C_F^2 - \frac{5}{8} C_F T n_f
        \,,\nonumber\\
        \beta_{02100}^e
        &=& -\frac{9}{8} 
        \,,\qquad%\nonumber\\
        \beta_{02010}^e
        \,\,=\,\, \frac{5}{4} 
        \,,\qquad%\nonumber\\
        \beta_{02001}^e
        \,\,=\,\, \frac{3}{4} 
        \,,\qquad%\nonumber\\
        \beta_{01200}^e
        \,\,=\,\, \frac{27}{64} 
        \,,\nonumber\\
        \beta_{01101}^e
        &=& -\frac{9}{16} 
        \,,\qquad%\nonumber\\
        \beta_{01020}^e
        \,\,=\,\, -\frac{15}{4}
        \,,\qquad%\nonumber\\
        \beta_{01002}^e
        \,\,=\,\, \frac{21}{32} 
        \,,
        \label{eq::betaDR}
\end{eqnarray}
where
\begin{equation}
\begin{split}
C_F &= \frac{N_c^2 -1}{2 N_c}\,,\qquad
C_A = N_c\,,\qquad
T=\frac{1}{2}
\end{split}
\end{equation}
are the usual colour factors of \qcd{}, and $n_f$ is the number of active
quark flavours.
In Eq.~(\ref{eq::defbetaDR}) we introduced five indices for the
coefficients of $\betaDRbar_s$ and $\beta_e$. However, we drop the last
three indices  
whenever there is no dependence on $\eta_r$. In particular,
$\beta_s$ depends on the $\eta_r$ only starting from 
four-loop order.
Note that those terms involving $\eta_r$ are only valid for 
$N_c = 3$, whereas the remaining ones hold for a general $SU(N_c)$ group.

As a first check on the results given in Eq.~(\ref{eq::betaDR}), we
specialize them to the supersymmetric Yang-Mills theory containing one
Majorana fermion in the adjoint representation, by setting $C_A = C_F = 2
T$, $n_f=1$, and $\asDRbar = \alpha_e = \eta_1$ and $\eta_2 = \eta_3 =
0$. Accordingly, we obtain for the non-vanishing coefficients of the
$\betaDRbar_s$ 
\begin{eqnarray}
 \betaDRbar_{20} =  \frac{3}{4}C_A \,, \quad 
 \betaDRbar_{30} =  \frac{3}{8}C_A^2 \,, \quad
 \betaDRbar_{40} =  \frac{21}{64}C_A^3
\,,
\end{eqnarray}
in agreement with Ref.~\cite{Avdeev:1981ew}.
Moreover, comparing these coefficients for pure QCD with the literature,
one finds that the two-loop 
result for $\betaDRbar_s$ and the one-loop result for $\beta_e$ agree
with Ref.~\cite{Jack:1993ws}. Actually, up to this order,
the result for the first two perturbative coefficients of $\beta_s$ is
the same in the \drbar{} and the \msbar{} scheme which is a well-known
consequence of
mass-independent renormalization schemes.  However, our three-loop
result for $\betaDRbar_s$ differs in the terms proportional to
$C_F^2T n_f$, $C_AC_FT n_f $ and $C_FT^2n_f^2$ from the one that can be
found in Ref.~\cite{Bern:2002zk}.

In order to explain this difference, let us have a closer look at the
method used in Ref.~\cite{Bern:2002zk}. The function $\betaDRbar_s$ 
was derived from the known result for $\betaMSbar_s$ by inserting
the relation between $\asDRbar$ and $\asMSbar$. 
The couplings $\alpha_e$ and $\asDRbar$, as well as 
their $\beta$-functions $\beta_e$ and $\betaDRbar_s$ were 
identified throughout the calculation. 
But as we will show shortly, this identification makes it impossible to
obtain consistent higher order results.
Keeping the couplings different, on the other hand, the
relation between $\asDRbar$ and $\asMSbar$ reads
\begin{eqnarray}
  \asDRbar &=& \asMSbar\left[1+\frac{\asMSbar}{\pi} \frac{C_A}{12}
  +\left(\frac{\asMSbar}{\pi}\right)^2
  \frac{11}{72}C_A^2 
  - \frac{\asMSbar}{\pi} \aepi
  \frac{1}{8} C_F T n_f
  + \ldots \right]
  \,,
  \label{eq::asMS2DR_2}
\end{eqnarray}
where the dots denote higher orders in $\asMSbar$, $\alpha_e$, and
$\eta_r$. We obtained this relation by noting that the value of
$\alpha_s$ in a physical renormalization scheme should not depend on
the regularization procedure:
\begin{equation}
\begin{split}
  \alpha_s^{\rm ph} = \left(z_s^{\rm ph,X}\right)^2 \alpha_s^{\rm X}\,,\qquad
  z_s^{\rm ph,X} &= Z_s^{\rm X}/Z_s^{\rm ph,X} \,,\qquad
  {\rm X} \in \{\overline {\rm MS},\overline {\rm DR}\}\\
  \Rightarrow
  \asDRbar &= \left(\frac 
	   {Z_s^{\rm ph,\overline{\rm DR}}\,Z_s^{\overline{\rm MS}}}
	   {Z_s^{\rm ph,\overline{\rm MS}}\,Z_s^{\overline{\rm DR}}}
	   \right)^2
  \,\asMSbar\,,
  \label{eq::asDRMSderiv}
\end{split}
\end{equation}
where $Z_s^{{\overline {\rm MS}}/{\overline {\rm DR}}}$ are the charge
renormalization constants using minimal
subtraction in \dreg{}/\dred{}, as defined above.  
For $Z_s^{{\rm ph},\overline{\rm MS}/\overline{\rm DR}}$,
on the other hand, we use \dreg{}/\dred{} combined with a {\it physical}
renormalization condition. We observe that the ratio in
Eq.\,(\ref{eq::asDRMSderiv}) is momentum independent, such that the
calculation amounts to keeping the constant finite pieces in the charge
renormalization constants
$Z_s^{{\rm ph},\overline{\rm MS}/\overline{\rm DR}}$.
Note that the various $Z_s$ in Eq.~(\ref{eq::asDRMSderiv}) 
depend on differently renormalized
$\alpha_s$, so that the equations have to be used iteratively at higher
orders of perturbation theory.

Equation~(\ref{eq::asMS2DR_2}) has to be inserted into
\begin{eqnarray}
  \betaDRbar_s(\asDRbar,\alpha_e,\{ \eta_r \})
  = \mu^2 \frac{{\rm d}}{{\rm d}\mu^2} \apiDR
  = \betaMSbar_s(\asMSbar) 
  \frac{\partial \asDRbar}{\partial \asMSbar} + 
  \beta_e(\asDRbar,\alpha_e,\{ \eta_r \}) 
  \frac{\partial \asDRbar}{\partial \alpha_e}+\ldots\,,
  \label{eq::beta_rel_2}
\end{eqnarray}
where the first equality is due to the definition of $\betaDRbar_s$ and
the second one is a consequence of the chain rule, with terms arising
through the $\eta_r$ represented by dots.  Using the three-loop
expression for $\betaMSbar_s$
(see~Refs.~\cite{vanRitbergen:1997va,Czakon:2004bu} and references
therein), we obtain the same result as in Eq.~(\ref{eq::betaDR}) which
  not only provides a powerful check on the various steps of the
  calculation, but also confirms the equivalence of \dreg{} and \dred{}
  at this order~\cite{Jack:1994bn}.

Let us stress that even if one sets $\alpha_e=\asDRbar$ in the final
result (cf.  Eq.~(\ref{eq::betaDR})), one does not arrive at the
expression for $\betaDRbar_s$ provided in Ref.~\cite{Bern:2002zk}.

Indeed, a way to see that the identification of $\asDRbar$ and
$\alpha_e$ at intermediate steps leads to inconsistent results is as
follows.  Whereas in the case of the $\beta$ function the error is a
finite, gauge parameter independent term, it leads to a much more
obvious problem for the quark mass renormalization: $Z_m$ will contain
non-local terms at three-loop order if $g_e=g_s$ is assumed throughout the
calculation, and $\gamma_m$ as evaluated from Eq.~(\ref{eq::Zm_gamma})
will not be finite.

%- }}}
%- {{{ Mass anomalous dimension:

\section{\label{sec::gamma}Mass anomalous dimension to three loops}

In this section we use the framework of
Section~\ref{sec::frame} in order to obtain 
the anomalous dimension of the quark masses within \dred{}
as defined in Eq.~(\ref{eq::Zm_gamma}).
The result will be derived both by a direct calculation of the relevant 
Feynman diagrams in \dred{}, as well as indirectly by using the result 
from \dreg{} and the \msbar{}--\drbar{} relation between 
the strong coupling and quark mass.

The evaluation of $Z_m$ to three-loop order proceeds along the same
lines as for the renormalization constants of the previous section.
However, in contrast to $Z_s$, the coupling $\alpha_e$ 
already appears at one-loop order. Thus the two-loop expression for 
$Z_e$ is required which can be obtained from Eq.~(\ref{eq::betaDR}).
At one-loop order we find complete agreement with the
result given in Ref.~\cite{Jack:1993ws}; the two-loop term is --- to our
knowledge --- new.

From the three-loop result for $Z_m$ we obtain the anomalous dimension
\begin{eqnarray}
\gammaDRbar_m(\asDRbar,\alpha_e, \{ \eta_r \})=
    - \sum_{i,j,k,l,m} \gammaDRbar_{ijklm} 
    \left(\apiDR\right)^i
    \left(\aepi\right)^j 
    \left(\frac{\eta_1}{\pi}\right)^k
    \left(\frac{\eta_2}{\pi}\right)^l 
    \left(\frac{\eta_3}{\pi}\right)^m 
\,,
\end{eqnarray}
with
\begin{eqnarray}
  \gammaDRbar_{10} &=& 
  \frac{3}{4}C_F  
        \,,\nonumber\\
  \gammaDRbar_{20} &=& 
  \frac{3}{32} C_F^2 + \frac{91}{96}  C_A C_F -
  \frac{5}{24} C_F T n_f
        \,,\nonumber\\
  \gammaDRbar_{11} &=& 
        - \frac{3}{8} C_F^2
        \,,\nonumber\\
  \gammaDRbar_{02} &=& 
        \frac{1}{4} C_F^2 -
  \frac{1}{8} C_A C_F + \frac{1}{8} C_F T n_f
        \,,\nonumber\\
  \gammaDRbar_{30} &=& 
        \frac{129}{128} C_F^3 -\frac{133}{256} C_F^2 C_A
        + \frac{10255}{6912} C_F C_A^2 
        + \frac{-23+24\zeta_3}{32} C_F^2 T n_f
        \nonumber\\&&
        -\left(\frac{281}{864} + \frac{3}{4}\zeta_3 \right) 
        C_A C_F T n_f
        - \frac{35}{432} C_F T^2 n_f^2
        \,,\nonumber\\ 
  \gammaDRbar_{21} &=& 
                - \frac{27}{64} C_F^3  
                - \frac{21}{32} C_F^2 C_A 
                - \frac{15}{256} C_F C_A^2 
                + \frac{9}{32} C_F^2 T n_f
                \,,\nonumber\\
  \gammaDRbar_{12} &=& 
  \frac{9}{8} C_F^3
  - \frac{21}{32} C_F^2 C_A 
        + \frac{3}{64} C_F C_A^2 
        + \frac{3}{64} C_A C_F T n_f 
        + \frac{3}{8} C_F^2 T n_f 
          \,,\nonumber\\
  \gammaDRbar_{03} &=& 
        - \frac{3}{8} C_F^3
  + \frac{3}{8} C_F^2 C_A 
  - \frac{3}{32} C_F C_A^2 
        + \frac{1}{8} C_A C_F T n_f 
        - \frac{5}{16} C_F^2 T n_f 
        - \frac{1}{32} C_F T^2 n_f^2 
        \,,\nonumber\\ 
   \gammaDRbar_{02100} &=& 
   \frac{3}{8} 
   \,,\qquad%\nonumber\\ 
   \gammaDRbar_{02010} \,\,=\,\, 
   -\frac{5}{12} 
   \,,\qquad%\nonumber\\ 
   \gammaDRbar_{02001} \,\,=\,\, 
   -\frac{1}{4} 
   \,,\qquad%\nonumber\\ 
   \gammaDRbar_{01200} \,\,=\,\, 
   -\frac{9}{64}    
   \,,\nonumber\\
   \gammaDRbar_{01020} &=&
   \frac{5}{4}  
   \,,\qquad%\nonumber\\
   \gammaDRbar_{01101} \,\,=\,\,
   \frac{3}{16} 
   \,,\qquad%\nonumber\\          
   \gammaDRbar_{01002} \,\,=\,\,   
    -\frac{7}{32} 
  \,.
  \label{eq::gamma_DRED}
\end{eqnarray}
Again the last three indices are suppressed whenever there is no
dependence on $\eta_r$.
Furthermore, those terms involving $\eta_r$ are only valid for 
$N_c = 3$, whereas the remaining ones hold for a general $SU(N_c)$ group.

On the other hand, $\gammaDRbar_m$ can be derived indirectly from the 
\msbar{} result obtained within \dreg{}. 
The analogous equation to~(\ref{eq::beta_rel_2}) is given by
\begin{eqnarray}
  \gammaDRbar_m(\asDRbar,\alpha_e,  \{ \eta_r \}) &=&
  \gammaMSbar_m
  \frac{\partial \ln \mDRbar}{\partial \ln \mMSbar}
  + \frac{\pi \betaMSbar_s}{\mDRbar} 
  \frac{\partial \mDRbar}{\partial \asMSbar}
  + \frac{\pi \beta_e}{\mDRbar}   
  \frac{\partial \mDRbar}{\partial \alpha_e}
  +\ldots
  \,,
  \label{eq::gamma_DRED_DREG}
\end{eqnarray}
which requires the two-loop relation between $\mDRbar$ and $\mMSbar$ in
order to obtain $\gammaDRbar_m$ to three loops.  The two-loop relation
between $\mDRbar$ and $\mMSbar$ can be computed in close analogy to
Eq.~(\ref{eq::asDRMSderiv}) by keeping not only the divergent but also
the finite parts in the calculation of the fermion propagator. Our
result reads
\begin{eqnarray}
  \mDRbar &=& \mMSbar\Bigg[1 -\aepi\frac{1}{4} C_F +
  \left(\apiMS\right)^2 \frac{11}{192} C_A C_F -\apiMS\aepi
  \left(\frac{1}{4} C_F^2 + \frac{3}{32}  C_A C_F \right) 
\nonumber\\
  &+ &   \left(\aepi\right)^2 \left( \frac{3}{32} C_F^2 
      +  \frac{1}{32} C_F T n_f\right)   
      + \ldots
\Bigg]
  \,,
  \label{eq::mDRmMS}
\end{eqnarray}
where the dots denote  higher orders in
$\asMSbar$, $\alpha_e$, and $\eta_r$.  The one- and two-loop terms of
Eq.~(\ref{eq::mDRmMS}) agree with Ref.~\cite{Avdeev:1997sz} in the
limit $\alpha_e=\asDRbar$.  Let us remark that in order to get
Eq.~(\ref{eq::mDRmMS}), also the one-loop relation between the \drbar{}
and \msbar{} version of the gauge parameter is a necessary ingredient
unless one works in Landau gauge ($\xi=1$).  We performed the
calculation for general covariant gauge and use the cancellation of the
gauge parameter in the final result as a welcome check.

As a further ingredient we need $\gammaMSbar_m$ which can be found
in Refs.~\cite{Chetyrkin:1997dh,Vermaseren:1997fq}.  Inserting this
result and Eq.~(\ref{eq::mDRmMS}) into~(\ref{eq::gamma_DRED_DREG}) leads
 to Eq.~(\ref{eq::gamma_DRED}). Again, this is a powerful check on
our calculation and shows the equivalence of \dred{} and \dreg{} at
this order. 
Note that in the indirect approach the $\eta_r$ enter only through
the factor $\beta_e$ in Eq.~(\ref{eq::gamma_DRED_DREG}).

The two-loop result of $\gammaDRbar_m$
can also be found in Ref.~\cite{Avdeev:1997sz} and we agree for
$\alpha_e=\alpha_s$.
The three-loop result for $\gammaDRbar_m$ is new.

The distinction between $\alpha_s$ and $\alpha_e$ in
Eq.~(\ref{eq::mDRmMS}) is essential for phenomenological analyses as can
be seen from the following numerical example. Assuming a supersymmetric
theory and integrating out the \susy{} particles at $\mu=M_Z$, we may
use $\asDRbar(M_Z)=\alpha_e(M_Z)=0.120$ as input, and then evolve
$\alpha_e$ and $\asDRbar$ separately to lower scales by using
Eqs.\,(\ref{eq::Zg_beta}), (\ref{eq::defbetaDR}), and
(\ref{eq::betaDR}). For $\mu_b=4.2$\,GeV, we arrive at\footnote{Only
one-loop running of $\alpha_s$ and $\alpha_e$ is applied here.}
$\asDRbar(\mu_b)= 0.218$ and $\alpha_e(\mu_b)= 0.167$, for
example. Using $m_b^{\overline{\rm MS}}(\mu_b)=4.2$\,GeV,
Eq.\,(\ref{eq::mDRmMS}) then leads to $m_b^{\overline{\rm
DR}}(\mu_b)=4.12$\,GeV.  If one wrongly identifies $\alpha_e$ with
$\asDRbar$ in Eq.\,(\ref{eq::mDRmMS}), one obtains a value for
$m_b^{\overline{\rm DR}}(\mu_b)$ which is roughly 30\,MeV smaller than
that. Note that this difference is of the same order of magnitude 
than the current uncertainty
on the $b$-quark mass
determination (see, e.g., Ref.~\cite{Kuhn:2001dm}).

Note that the identification $\alpha_e=\alpha_s$ has also been made in
Eq.~(26) of Ref.~\cite{Aguilar-Saavedra:2005pw} for $\mu=M_Z$ 
(see also Ref.~\cite{Baer:2002ek}), which incorporates 
our Eq.~(\ref{eq::mDRmMS}) for $n_f=5$.
This induces an inconsistency of order $\alpha_s^2(M_Z)$, whose numerical
effect is quite small.

%- }}}
%- {{{ Conclusions:

\section{\label{sec::concl}Conclusions}

In many cases, \dred{} poses an attractive alternative to \dreg{} ---
not only for supersymmetric theories. We computed the \qcd{}
renormalization group function of the strong 
coupling constant ($\beta$) and of the quark masses ($\gamma_m$) to
three-loop order in this scheme using two different methods.
The agreement of the results obtained in both ways confirms
the equivalence of the \drbar{} and the \msbar{} renormalization scheme
at this order, in the sense that they are related by an analytic
redefinition of the couplings and masses~\cite{Jack:1994bn}.
Furthermore, we find that the three-loop $\beta$-function found in the
literature differs from ours. We trace this difference to the fact that
the evanescent coupling of the $q\bar q\varepsilon$ vertex had been
identified wrongly with $\alpha_s$ in Ref.\,\cite{Bern:2002zk}.

Let us stress that higher order calculations within the framework of
\dred{} should also be useful in the context of the Minimal
Supersymmetric Standard Model where precision calculations will be
important in order to be prepared for measurements at the CERN Large
Hadron Collider and other future high energy experiments.

%- }}}
%- {{{ Acknowledgements:

\vspace*{1em}

\noindent
{\large\bf Acknowledgements}\\ We would like to thank K.G.~Chetyrkin and
D.R.T.~Jones for 
carefully reading the manuscript and many useful comments, as well as
Z.~Bern  for discussions and comments.  This work was
supported by the DFG through SFB/TR~9. RH is supported by the DFG, {\it
Emmy Noether program}.

%- }}}
%- {{{ bibliography:

%- }}}


\begin{thebibliography}{99}

%
% beta3l_ref.tex -- generated by sortref-2.3.5  
% ((C) R. Harlander, http://www.robert-harlander.de/software/)
% on Thu Aug 24 15:56:05 CEST 2006
%

%1
\bibitem{'tHooft:1972fi}
  G.~'t Hooft and M.~J.~G.~Veltman,
  Nucl.\ Phys.\ B {\bf 44} (1972) 189.
  %%CITATION = NUPHA,B44,189;%%

%2
\bibitem{Bollini:1972ui}
  C.~G.~Bollini and J.~J.~Giambiagi,
  Nuovo Cim.\ B {\bf 12} (1972) 20.
  %%CITATION = NUCIA,B12,20;%%

%3
\bibitem{DRED}
W.~Siegel,
Phys.\ Lett.\ B {\bf 84} (1979) 193.
%%CITATION = PHLTA,B84,193;%%

%4
\bibitem{Siegel:1980qs}
  W.~Siegel,
  Phys.\ Lett.\ B {\bf 94} (1980) 37.
  %%CITATION = PHLTA,B94,37;%%

%5
\bibitem{Avdeev:1981vf}
  L.~V.~Avdeev, G.~A.~Chochia and A.~A.~Vladimirov,
  Phys.\ Lett.\ B {\bf 105} (1981) 272.
  %%CITATION = PHLTA,B105,272;%%

%6
\bibitem{Stockinger:2005gx}
  D.~St\"ockinger,
  JHEP {\bf 0503} (2005) 076
  [hep-ph/0503129].
  %%CITATION = HEP-PH 0503129;%%

%7
\bibitem{Hollik:2005nn}
  W.~Hollik and D.~St\"ockinger,
  Phys.\ Lett.\ B {\bf 634} (2006) 63
  [hep-ph/0509298].
  %%CITATION = HEP-PH 0509298;%%

%8
\bibitem{Jack:1993ws}
  I.~Jack, D.~R.~T.~Jones and K.~L.~Roberts,
  Z.\ Phys.\ C {\bf 62} (1994) 161
  [hep-ph/9310301].
  %%CITATION = HEP-PH 9310301;%%

%9
\bibitem{Smith:2004ck}
  J.~Smith and W.~L.~van Neerven,
  Eur.\ Phys.\ J.\ C {\bf 40} (2005) 199
  [hep-ph/0411357].
  %%CITATION = HEP-PH 0411357;%%

%10
\bibitem{Signer:2005iu}
  A.~Signer and D.~St\"ockinger,
  Phys.\ Lett.\ B {\bf 626} (2005) 127
  [hep-ph/0508203].
  %%CITATION = HEP-PH 0508203;%%

%11
\bibitem{Kunszt:1993sd}
  Z.~Kunszt, A.~Signer and Z.~Trocsanyi,
  Nucl.\ Phys.\ B {\bf 411} (1994) 397
  [hep-ph/9305239].
  %%CITATION = HEP-PH 9305239;%%

%12
\bibitem{Bern:2002zk}
  Z.~Bern, A.~De Freitas, L.~J.~Dixon and H.~L.~Wong,
  Phys.\ Rev.\ D {\bf 66} (2002) 085002
  [hep-ph/0202271].
  %%CITATION = HEP-PH 0202271;%%

%13
\bibitem{Capper:1979ns}
  D.~M.~Capper, D.~R.~T.~Jones and P.~van Nieuwenhuizen,
  Nucl.\ Phys.\ B {\bf 167} (1980) 479.
  %%CITATION = NUPHA,B167,479;%%

%14
\bibitem{Bednyakov:2002sf}
  A.~Bednyakov, A.~Onishchenko, V.~Velizhanin and O.~Veretin,
  Eur.\ Phys.\ J.\ C {\bf 29} (2003) 87
  [hep-ph/0210258].
  %%CITATION = HEP-PH 0210258;%%

%15
\bibitem{Jack:1994kd}
  I.~Jack and D.~R.~T.~Jones,
  Phys.\ Lett.\ B {\bf 333} (1994) 372
  [hep-ph/9405233].
  %%CITATION = HEP-PH 9405233;%%

%16
\bibitem{Jack:1994rk}
  I.~Jack, D.~R.~T.~Jones, S.~P.~Martin, M.~T.~Vaughn and Y.~Yamada,
  Phys.\ Rev.\ D {\bf 50} (1994) 5481
  [hep-ph/9407291].
  %%CITATION = HEP-PH 9407291;%%

%17
\bibitem{Bardeen:1978yd}
  W.~A.~Bardeen, A.~J.~Buras, D.~W.~Duke and T.~Muta,
  Phys.\ Rev.\ D {\bf 18} (1978) 3998.
  %%CITATION = PHRVA,D18,3998;%%

%18
\bibitem{vanRitbergen:1997va}
  T.~van Ritbergen, J.~A.~M.~Vermaseren and S.~A.~Larin,
  Phys.\ Lett.\ B {\bf 400} (1997) 379
  [hep-ph/9701390].
  %%CITATION = HEP-PH 9701390;%%

%19
\bibitem{Czakon:2004bu}
  M.~Czakon,
  Nucl.\ Phys.\ B {\bf 710} (2005) 485
  [hep-ph/0411261].
  %%CITATION = HEP-PH 0411261;%%

%20
\bibitem{Chetyrkin:1997dh}
  K.~G.~Chetyrkin,
  Phys.\ Lett.\ B {\bf 404} (1997) 161
  [hep-ph/9703278].
  %%CITATION = HEP-PH 9703278;%%

%21
\bibitem{Vermaseren:1997fq}
  J.~A.~M.~Vermaseren, S.~A.~Larin and T.~van Ritbergen,
  Phys.\ Lett.\ B {\bf 405} (1997) 327
  [hep-ph/9703284].
  %%CITATION = HEP-PH 9703284;%%

%22
\bibitem{Harlander:2005wm}
  R.~Harlander, L.~Mihaila and M.~Steinhauser,
  Phys.\ Rev.\ D {\bf 72} (2005) 095009
  [hep-ph/0509048].
  %%CITATION = HEP-PH 0509048;%%

%23
\bibitem{Steinhauser:1998cm}
  M.~Steinhauser,
  Phys.\ Rev.\ D {\bf 59} (1999) 054005
  [hep-ph/9809507].
  %%CITATION = HEP-PH 9809507;%%

%24
\bibitem{Nogueira:1991ex}
  P.~Nogueira,
  J.\ Comput.\ Phys.\  {\bf 105} (1993) 279.
  %%CITATION = JCTPA,105,279;%%

%25
\bibitem{Seidensticker:1999bb}
  T.~Seidensticker,
  [hep-ph/9905298].
  %%CITATION = HEP-PH 9905298;%%

%26
\bibitem{Harlander:1997zb}
  R.~Harlander, T.~Seidensticker and M.~Steinhauser,
  Phys.\ Lett.\ B {\bf 426} (1998) 125
  [hep-ph/9712228].
  %%CITATION = HEP-PH 9712228;%%

%27
\bibitem{Larin:1991fz}
  S.~A.~Larin, F.~V.~Tkachov and J.~A.~M.~Vermaseren,
NIKHEF-H-91-18

%28
\bibitem{Avdeev:1981ew}
  L.~V.~Avdeev and O.~V.~Tarasov,
  Phys.\ Lett.\ B {\bf 112} (1982) 356.
  %%CITATION = PHLTA,B112,356;%%

%29
\bibitem{Jack:1994bn}
  I.~Jack, D.~R.~T.~Jones and K.~L.~Roberts,
  Z.\ Phys.\ C {\bf 63} (1994) 151
  [hep-ph/9401349].
  %%CITATION = HEP-PH 9401349;%%

%30
\bibitem{Avdeev:1997sz}
  L.~V.~Avdeev and M.~Y.~Kalmykov,
  Nucl.\ Phys.\ B {\bf 502} (1997) 419
  [hep-ph/9701308].
  %%CITATION = HEP-PH 9701308;%%

%31
\bibitem{Kuhn:2001dm}
  J.~H.~K\"uhn and M.~Steinhauser,
  Nucl.\ Phys.\ B {\bf 619} (2001) 588
  [Erratum-ibid.\ B {\bf 640} (2002) 415]
  [arXiv:hep-ph/0109084].
  %%CITATION = HEP-PH 0109084;%%

%32
\bibitem{Aguilar-Saavedra:2005pw}
  J.~A.~Aguilar-Saavedra {\it et al.},
  Eur.\ Phys.\ J.\ C {\bf 46} (2006) 43
  [hep-ph/0511344].
  %%CITATION = HEP-PH 0511344;%%

%\cite{Baer:2002ek}
\bibitem{Baer:2002ek}
  H.~Baer, J.~Ferrandis, K.~Melnikov and X.~Tata,
  %``Relating bottom quark mass in DR-bar and MS-bar regularization
  %  schemes,''
  Phys.\ Rev.\ D {\bf 66} (2002) 074007
  [arXiv:hep-ph/0207126].
  %%CITATION = HEP-PH 0207126;%%

\end{thebibliography}
\end{document}